\documentclass{iaus}
\usepackage{graphicx}
\usepackage{amsmath}
\usepackage[square]{natbib}
\bibpunct{[}{]}{,}{n}{}{,} 

\newcommand{\aanda}{\textit{A\&A}}

\newcommand{\aj}{\textit{AJ}}
\newcommand{\apj}{\textit{ApJ}}

\newcommand{\mnras}{\textit{MNRAS}}

\newcommand{\pasp}{\textit{PASP}}

\title[RR Lyrae stars in the {\sc sdss} system] 
{Physical parameters for RR Lyrae stars\\ in the {\sc sdss} filter system}

\author[M. Catelan et al.]   
{
M\'arcio Catelan,$^{1,2}$
Gabriel I. Torrealba,$^{1,2}$
Claudio C\'aceres,$^{1,2,3}$\\
Horace A. Smith,$^{4}$ 
Nathan De Lee,$^{5}$ 
\and 
Michael Fitzgerald$^{6}$
}

\affiliation{
$^1$Departamento de Astronom{\'\i}a y Astrof{\'\i}sica, Pontificia Universidad Cat{\'o}lica de Chile,\\
Av.\ Vicu{\~n}a Mackenna 4860, 782-0436 Macul, Santiago, Chile  \\ email: {\tt mcatelan@astro.puc.cl} \\[\affilskip]
$^2$The Milky Way Millennium Nucleus, Santiago, Chile  \\[\affilskip]
$^3$Departamento de F\'{\i}sica y Astronom\'{\i}a, Universidad de Valpara\'{\i}so, 
Av. Gran Breta\~{n}a 1111, Playa Ancha, Valpara\'{\i}so, Chile \\[\affilskip]
$^4$Department of Physics and Astronomy, Michigan State University, East Lansing, MI 48824, USA\\[\affilskip]
$^5$Department of Physics and Astronomy, Vanderbilt University, Nashville, TN 37235, USA\\[\affilskip]
$^6$Department of Physics \& Astronomy, Macquarie University, NSW 2109, Australia 
}

\pubyear{2012}
\volume{289}  
\pagerange{126--129}
\jname{Advancing the Physics of Cosmic Distances}
\editors{Richard de Grijs, ed.}
\begin{document}

\maketitle

\begin{abstract}
We present a calibration of the metallicity and physical parameters
(temperature, luminosity, gravity, mass, radius) for RR Lyrae stars
using the $ugriz$ {\sc sdss} photometric system. Our work is based on
calculations of synthetic horizontal branches (HBs), fully taking into
account evolutionary effects for a wide range in metallicities and HB
morphologies. We provide analytical fits that are able to provide all
quantities mentioned with very high (internal) precision, based
solely on mean {\sc sdss} magnitudes and colors.  
\end{abstract}

\firstsection 
\section{Introduction}

The Sloan Digital Sky Survey ({\sc sdss}; York et al. 2000; Aihara et
al. 2011) represents the dawn of a new era in astronomy, in which
wide-field sky surveys play increasingly important roles. Such surveys
are of key importance for our understanding of structure in the
Universe, both locally and at cosmological distances. In this sense,
it is particularly important that the properties of key distance
indicators, such as RR Lyrae stars, are well-understood and properly
calibrated on the basis of {\sc sdss} photometric indices.

In the wake of the {\sc sdss} survey itself, the {\sc sdss}
photometric system (Fukugita et al. 1996, 2011) has gained much
visibility, and {\sc sdss} filters are now commonly available at all
major observatories. Indeed, many of the current and future wide-field
dedicated telescopes and surveys, including the Large Synoptic Survey
Telescope ({\sl LSST}; Ivezi\'c et al. 2008b), {\sl PanSTARRS} (Kaiser
et al. 2002; Stubbs et al. 2007), the {\sl VLT} Survey Telescope ({\sl
VST}; Kuijken et al. 2002), the Dark Energy Survey ({\sc des}; Tucker
et al. 2007), and {\sl SkyMapper} (Keller et al. 2007; Bessell et
al. 2011), will be (or are already) carried out using filter systems
that generally bear close resemblance to the original {\sc sdss}
system.

While it is known that the {\sc sdss} system possesses great
scientific potential for a variety of science applications, including
stellar populations (e.g., Lenz et al. 1998; Helmi et al. 2003;
Ivezi\'c et al. 2008a; Lardo et al. 2011; Vickers et al. 2012), the
behavior of variable stars in general in such filter systems has not
yet been as extensively studied as for more traditional filter
systems, particularly the Johnson--Cousins system. For instance, no
systematic studies of RR Lyrae variability in globular clusters has
yet been carried out in the {\sc sdss} system, and similarly
theoretical analyses of RR Lyrae variable stars in the {\sc sdss}
system are almost entirely lacking, with the studies of Marconi et
al. (2006) and C\'aceres \& Catelan (2008) seemingly representing the
sole exceptions. To extract the maximum amount of information from
extensive RR Lyrae databases that are increasingly becoming available
in the {\sc sdss} (or similar) systems (e.g., Sesar 2011; Sesar et
al. 2007, 2010, 2011), more extensive theoretical analyses are clearly
needed.

In this sense, we recently started a systematic study, based on
theoretical models and synthetic calculations for horizontal-branch
(HB) stars, to define precise relations that should allow one to
calculate distances, reddening and metallicity values, and physical
parameters of RR Lyrae stars from {\sc sdss} photometric observations.
In C\'aceres \& Catelan (2008) we presented the first detailed
calibration of the RR Lyrae period--luminosity (PL) and period--color
(PC) relations in the {\sc sdss} system.

One shortcoming of the C\'aceres \& Catelan (2008) PL and PC
calibrations is that they require {\it a priori} knowledge of the
metallicity, which is frequently not available, especially for field
stars. Currently this requires either spectroscopic information or
Fourier decomposition of $V$-band light curves, whose parameters have
been calibrated in terms of metallicity (Jurcsik \& Kov\'acs 1996;
Morgan et al. 2007). However, as pointed out by Jurcsik \& Kov\'acs
(1996), such calibrations of Fourier-decomposition parameters are not
applicable to all RR Lyrae stars, requiring very well-behaved light
curves. In addition, exceedingly complete phase coverage is required
for the computation of reliable Fourier parameters. Therefore,
calibrations that are based solely on the {\em average} photometric
properties of the RR Lyrae stars are certainly desired.

The main purpose of the present project is thus to extend the
C\'aceres \& Catelan (2008) study, by providing an additional set of
analytical expressions that allow one to compute metallicities,
luminosities, temperatures, masses, gravities, and radii of RR Lyrae
stars, solely on the basis of their average photometric
properties. Our numerical experiments have shown that it is possible
to derive fairly precise relations in the multiband {\sc sdss} and
Str\"omgren (1963) systems, whereas we have not achieved similar
success for the Johnson--Cousins $UBVRI$ system.

\section{Models}

The HB simulations computed in the present paper follow the same
techniques as described in Catelan (2004) and Catelan et al. (2004),
to which the reader is referred for further details and references
about the HB synthesis method. In the present paper, we use the same
HB simulations as already employed in our study of the PL and PC
relations in the {\sc sdss} filter system (C\'aceres \& Catelan 2008).

\section{Pseudocolors for the {\sc sdss} System}

In this study, we define two {\em pseudocolors} in the {\sc sdss}
system,
\begin{eqnarray}
C_{0} & = & (u-g)_{0}-(g-r)_{0}; \\
m_{0} & = & (g-r)_{0}-(r-i)_{0}. 
\label{eq:m0c0}
\end{eqnarray}
These indices are patterned after the well-known gravity and
metallicity indices of the Str\"omgren (1963) system. Similarly to
what happens in the latter system, both indices are fairly insensitive
to reddening, with $E(C_1)\approx -0.32 \, E(B-V)$ and $E(m_1) \approx
-0.38$ \hbox{$E(B-V)$}. C\'aceres \& Catelan (2008) already showed
that the introduction of $C_0$ allows one to compute precise
relationships for the derivations of absolute magnitudes and colors in
the {\sc sdss} system; we have found that $m_0$ is also very helpful,
as far as derivation of metallicity and the remaining physical
parameters of RR Lyrae stars is concerned.

\section{Results}

Full details of this investigation will be provided elsewhere; here we
give only a brief preview of our results, showing a preliminary, small
subsample of the relations derived so far. In all relations that
follow, the (fundamentalized) period is given in days.

\subsection{Temperature}

\begin{equation}
\log T_{\rm eff}  =  \mathcal{A} + \mathcal{B} (g-r)_0 + \mathcal{C} m_{0} + \mathcal{D} m_{0}^{2}
                     + \mathcal{E} (g-r)_0 (\log P)^2,
\label{eq:TEFF}
\end{equation}
where $\mathcal{A} = 3.84493$, $\mathcal{B} = -0.43149$, $\mathcal{C}
= 0.24311$, $\mathcal{D} = 0.21078$, and $\mathcal{E} = 0.15544$. The
correlation coefficient is $\mathcal{R} = 0.99997$, and the standard
error of the estimate amounts to $\sigma = 1.7 \times 10^{-4}$~dex.

\subsection{Gravity}

\begin{equation}
\log g  =  \mathcal{A} + \mathcal{B} \log P + \mathcal{C} m_0 + \mathcal{D} (g-r)_0,  
\label{eq:LOGG}
\end{equation}
where $\mathcal{A} = 2.49342$, $\mathcal{B} = -1.22603$, $\mathcal{C}
= -0.36479$, and $\mathcal{D} = 0.24119$. The correlation coefficient
is $\mathcal{R} = 0.99995$, and the standard error of the estimate is
$\sigma = 0.0011$~dex.

\subsection{Luminosity}

\begin{eqnarray}
\log\left(\frac{L}{{\rm L}_{\odot}}\right)   &=&  \mathcal{A} + \mathcal{B} M_g^2 \log P + \mathcal{C} M_g + \mathcal{D} (g-r)_0\nonumber  \\
                                       &+&  \mathcal{E} \log P + \mathcal{F} \log P (g-r)_0 + \mathcal{G} m_0^2, 
\label{eq:LOGL}
\end{eqnarray}
where $\mathcal{A} = 1.78262$, $\mathcal{B} = 0.07619$, $\mathcal{C} =
-0.49604$, $\mathcal{D} = 1.05627$, $\mathcal{E} = -0.40866$,
$\mathcal{F} = 0.37410$, $\mathcal{G} = -1.21339$, and $M_g$ can be
obtained from the equations provided in C\'aceres \& Catelan
(2008). The correlation coefficient here is $\mathcal{R} = 0.99945$,
and the standard error of the estimate is $\sigma = 0.0019$~dex.

\subsection{Mass and Radius}

Note that, by using the previous equations, the definition of surface
gravity, and the Stefan--Boltzmann law, one can also derive
high-precision masses and radii. More specifically, masses derived on
the basis of these expressions have an internal $\sigma = 0.0017$
M$_{\odot}$, whereas in the case of radii one has $\sigma = 0.0026$
R$_{\odot}$.

\subsection{Metallicity}

\begin{eqnarray}
{\rm [Fe/H]}  = \mathcal{A} + \mathcal{B} \frac{{\rm [Fe/H]}_{0}^{2}}{(u-g)_0} + \mathcal{C} m_0 (\log P)^2
                        + \mathcal{D} m_0 {\rm [Fe/H]}_{0}^{2}, 
\label{eq:METAL}
\end{eqnarray}
\noindent with 
\begin{eqnarray}
{\rm [Fe/H]_{0}}  = \mathcal{E} + \mathcal{F} C_0^2 m_{0}, 
\label{eq:METAL0}
\end{eqnarray}
where $\mathcal{A} = -0.78752$, $\mathcal{B} = -0.52606$, $\mathcal{C}
= 20.20356$, $\mathcal{D} = 0.61526$, $\mathcal{E} = -3.29763$, and
$\mathcal{F} = 13.98842$. The correlation coefficient is $\mathcal{R}
= 0.99445$, and the standard error of the estimate is $\sigma =
0.039$~dex.

\subsection{Robustness of the Relations}

We have run extensive tests to check the robustness of the relations
provided, based on independent sets of HB models covering a wider
range of metallicities than originally used in C\'aceres \& Catelan
(2008), as well as empirical data, particularly for [Fe/H] (e.g., De
Lee et al. 2007; De Lee 2008; and references therein). We find that
the relations can be safely used, even outside their original range of
applicability, provided that errors in the measurements of the colors
and magnitudes of the `equivalent static star' are not larger than
0.01--0.05~mag. In this context, work along the lines of Bono et al.
(1995), but specifically focused on the {\sc sdss} filter system, is
strongly encouraged.

\acknowledgements This work is supported by the Chilean Ministry for
the Economy, Development, and Tourism's Programa Iniciativa
Cient\'{\i}fica Milenio (grant P07-021-F, awarded to The Milky Way
Millennium Nucleus), by Proyecto Fondecyt Regular \#1110326, by the
Center for Astrophysics and Associated Technologies (PFB-06), and by
Proyecto Anillo ACT-86. CC acknowledges support from ALMA-CONICYT
project 31100025. HAS and NDL thank the US National Science Foundation
for support under grant AST0707756.


\vspace{-0.4cm}
\begin{thebibliography}{}
\bibitem[Aihara et al.(2011)]{haea11} Aihara, H., Allende Prieto, C.,
  An, D., et al.\ 2011, {\it ApJS}, 193, 29; erratum: {\it ApJS}, 195,
  26

\bibitem[Bessell et al.(2011)]{mbea11} Bessell, M., Bloxham, G.,
  Schmidt, B., et al.\ 2011, \pasp, 123, 789

\bibitem[Bono et al.(1995)]{gbea95} Bono, G., Caputo, F., \&
  Stellingwerf, R.F.\ 1995, {\it ApJS}, 99, 263

\bibitem[C\'aceres \& Catelan(2008)]{ccmc08} C\'aceres, C., \&
  Catelan, M. 2008, {\it ApJS}, 179, 242

\bibitem[Catelan(2004)]{mc04} Catelan, M.\ 2004, \apj, 600, 409

\bibitem[Catelan et al.(2004)Catelan, Pritzl, \& Smith]{mcea04}
  Catelan, M., Pritzl, B.J., \& Smith, H.A.\ 2004, {\it ApJS}, 154,
  633

\bibitem[De Lee(2008)]{ndl08} De Lee, N. 2008, Ph.D. Thesis, Michigan
  State University, USA

\bibitem[De Lee, Smith, \& Beers(2007)]{dsb07} De Lee, N.M., Smith,
  H.A., \& Beers, T.C. 2007, {\em Bull. Am. Astron. Soc.}, 39, 776

\bibitem[Fukugita et al.(1996)]{mfea96} Fukugita, M., Ichikawa, T.,
  Gunn, J.E., et al.\ 1996, \aj, 111, 1748
  
\bibitem[Fukugita et al.(2011)]{mfea11} Fukugita, M., Yasuda, N., Doi,
  M., Gunn, J.E., \& York, D.G.\ 2011, \aj, 141, 47
  
\bibitem[Helmi et al.(2003)]{ahea03} Helmi, A., Ivezi{\'c}, {\v Z}.,
  Prada, F., et al.\ 2003, \apj, 586, 195

\bibitem[Ivezi{\'c} et al.(2008a)]{ziea08a} Ivezi{\'c}, {\v Z}.,
  Sesar, B., Juri{\'c}, M., et al.\ 2008a, \apj, 684, 287
	
\bibitem[Ivezic et al.(2008b)]{ziea08b} Ivezi\'{c}, {\v Z}, Tyson,
  J.~A., Acosta, E., et al.\ 2008b, {\it LSST: from Science Drivers to
    Reference Design and Anticipated Data Products} (arXiv:0805.2366)

\bibitem[Jurcsik \& Kov\'acs(1996)]{jk96} Jurcsik, J., \& Kov\'acs,
  G. 1996, \aanda, 312, 111

\bibitem[Kaiser et al.(2002)]{nkea02} Kaiser, N., Aussel, H., Burke,
  B.E., et al.\ 2002, {\it Proc. SPIE}, 4836, 154
  
\bibitem[Keller et al.(2007)]{skea07} Keller, S.C., Schmidt, B.P.,
  Bessell, M.S., et al.\ 2007, {\it Publ. Astron. Soc. Aus.}, 24, 1
  
\bibitem[Kuijken et al.(2002)]{kkea02} Kuijken, K., Bender, R.,
  Cappellaro, E., et al.\ 2002, {\it The Messenger}, 110, 15

\bibitem[Lardo et al.(2011)]{clea11} Lardo, C., Bellazzini, M.,
  Pancino, E., Carretta, E., Bragaglia, A., \& Dalessandro, E. 2011,
  \aanda, 525, A114

\bibitem[Lenz et al.(1998)]{dlea98} Lenz, D.D., Newberg, J., Rosner,
  R., Richards, G.T., \& Stoughton, C.\ 1998, {\it ApJS}, 119, 121
  
\bibitem[Marconi et al.(2006)]{mmea06} Marconi, M., Cignoni, M., Di
  Criscienzo, M., et al.
2006, \mnras, 371, 1503

\bibitem[Morgan et al.(2007)Morgan, Wahl, \& Wieckhorst]{smea07}
  Morgan, S.M., Wahl, J.N., \& Wieckhorst, R.M. 2007, \mnras, 374,
  1421

\bibitem[Sesar(2011)]{bs11} Sesar, B.\ 2011, {\it Carnegie
  Obs. Astrophy. Ser.}, 5, 135
  
\bibitem[Sesar et al.(2010)]{bsea10} Sesar, B., Ivezi{\'c}, {\v Z}.,
  Grammer, S.H., et al.\ 2010, \apj, 708, 717

\bibitem[Sesar et al.(2007)]{bsea07} Sesar, B., Ivezi{\'c}, {\v Z}.,
  Lupton, R.H., et al.\ 2007, \aj, 134, 2236

\bibitem[Sesar et al.(2011)]{bsea11} Sesar, B., Stuart, J.S.,
  Ivezi{\'c}, {\v Z}., et al.\ 2011, \aj, 142, 190

\bibitem[Str{\"o}mgren(1963)]{bs63} Str{\"o}mgren, B.\ 1963, {\it
  QJRAS}, 4, 8

\bibitem[Stubbs et al.(2007)]{csea07} Stubbs, C.W., High, F.W.,
  George, M.R., et al.\ 2007, \pasp, 119, 1163
  
\bibitem[Tucker et al.(2007)]{dtea07} Tucker, D.L., Annis, J.T., Lin,
  H., et al.\ 2007, in: {\it The Future of Photometric,
    Spectrophotometric and Polarimetric Standardization} (Sterken, C.,
  ed.), ASP Conf. Ser., 364, 187
  
\bibitem[Vickers et al.(2012)Vickers, Grebel, \& Huxor]{jvea12}
  Vickers, J.J., Grebel, E.K., \& Huxor, A.P.\ 2012, \aj, 143, 86
	
\bibitem[York et al.(2000)]{dyea00} York, D.G., Adelman, J.,
  Anderson, J.E., Jr., et al.\ 2000, \aj, 120, 1579

\end{thebibliography}
\end{document}